\documentclass[a4paper,12pt]{spieman}  
\usepackage{amsmath,amsfonts,amssymb}
\usepackage{graphicx}
\usepackage{setspace}
\usepackage{tocloft}

\title{Rigorous vector wave propagation for arbitrary flat media}

\author[a,*]{Steven P. Bos}
\author[a]{Sebastiaan Y. Haffert }
\author[a]{Christoph U. Keller}
\affil[a]{Leiden Observatory, Leiden University, P.O. Box 9513, 2300 RA Leiden, The Netherlands}

\renewcommand{\Im}{\operatorname{Im}}

\cftpagenumbersoff{figure}
\cftpagenumbersoff{table} 
\begin{document} 
\maketitle

\begin{abstract}
Precise modelling of the (off-axis) point spread function (PSF) to identify geometrical and polarization aberrations is important for many optical systems. In order to characterise the PSF of the system in all Stokes parameters, an end-to-end simulation of the system has to be performed in which Maxwell's equations are rigorously solved. We present the first results of a python code that we are developing to perform multiscale end-to-end wave propagation simulations that include all relevant physics. Currently we can handle plane-parallel near- and far-field vector diffraction effects of propagating waves in homogeneous isotropic and anisotropic materials, refraction and reflection of flat parallel surfaces, interference effects in thin films and unpolarized light. We show that the code has a numerical precision on the order of $\sim10^{-16}$ for non-absorbing isotropic and anisotropic materials. For absorbing materials the precision is on the order of $\sim10^{-8}$. The capabilities of the code are demonstrated by simulating a converging beam reflecting from a flat aluminium mirror at normal incidence.
\end{abstract}

\keywords{Vector Wave Propagation, Polarization, Polarization Aberrations, Simulations, Modal Methods}

{\noindent \footnotesize\textbf{*} \linkable{stevenbos@strw.leidenuniv.nl} }
\section{Introduction}\label{sect:introduction}  
The precise characterization of optical systems is important for many applications, e.g. lithography\cite{melville2011computational}, microscopy\cite{hansen1988overcoming} and astronomy\cite{breckinridge2015polarization}. Technology pushing the boundaries of the accuracy and sensitivity of optical systems, requires simulation tools that predict to a high accuracy the point spread function (PSF) of the systems.\\ 
\subsection{Polarization aberrations}
The monochromatic scalar aberrations of the wavefront, i.e. the phase and amplitude deviations from the ideal wavefront, were for a long time the only aberrations taken into account when simulating optical systems. These aberrations are caused by variations in the optical path length (OPL) and transmission for rays propagating through different parts of the optical system. These and chromatic aberrations have the the largest impact on image formation. As optical systems start to operate under more extreme numerical apertures and host more sensitive polarimeters, aberrations that affect the polarization state, which are much weaker than scalar aberrations, become relevant. We define polarization aberrations as the undesired deviations from the ideal system, a system which preserves the polarization state of light. Similar to scalar aberrations, polarization aberrations are caused by amplitude and phase effects. Diattenuation is the difference in transmission between two orthogonal polarization states. This can be due to different amplitude effects for the polarization states when reflecting off a surface. Retardance is the difference in OPL between two orthogonal polarization states, which is for example the result of propagation in anisotropic materials.\\
\subsection{Simulation methods}
High accuracy optical simulation codes need to be able to perform end-to-end simulations of systems with meter to wavelength size components and have all relevant physics in their arsenal: rigorous vector diffraction that can handle isotropic and anisotropic (in)homogeneous materials, arbitrarily shaped surfaces, reflection and transmission of these surfaces, interference effects in thin films and wave plates, and unpolarized light. Many different methods exists for simulating optical systems. We briefly discuss three common methods that include the vector nature of light.\\
\indent Finite-difference time-domain (FDTD) methods solve Maxwell's equations in the time domain on a fine grid that includes the entire optical system. They approximate derivatives in Maxwell's equations by numerical methods as finite differences. The entire grid is updated every time step to evaluate the electromagnetic field in every point in the grid. These methods require fine temporal and spatial discretisation on the wavelength scale, which in the case of ELT's result in extremely large computational domains. The properties of materials at any position in the grid can be specified and therefore a broad range of structures can be modelled. FDTD operates in the time domain and thus the response of the system across a wide frequency range can be modelled in one simulation.\\
\indent Polarization ray tracing codes, such as Polaris-M\cite{chipman2015polaris} and the polarization ray tracer of Code V,  keep track of a `local' Jones matrix for arbitrary rays propagating through the system. The Jones matrix keeps track of all polarization changes due to diattenuation and retardance effects. Performing this for all rays that form the pupil of a optical system generates a Jones pupil\cite{ruoff2009orientation}. The Jones pupil includes polarization effects, but also geometrical phase effects that are irrespective of polarization. By taking the Fourier transform of all individual elements of the Jones pupil, a generalised form of the PSF is obtained.\\ 
\indent The modal method (MM) decomposes the electromagnetic field on global eigenmodes in the spatial and temporal frequency domain. Reducing Maxwell's differential equations to algebraic equations that have a set of eigenmodes as fully vectorial rigorous solutions. The system is divided into subdomains with invariant refractive index profiles along the propagation axis of the propagating electromagnetic waves. Each subdomain has a specific set of eigenmodes that form the basis on which the electromagnetic field is expanded, these modes can be plane waves for homogeneous materials or have more complicated profiles in inhomogeneous structures. Each eigenmode preserves its mode profile when propagating through the subdomain and therefore can be propagated an arbitrary distance within one operation. The discretization along the axis of propagation of MM's is therefore directly related to the amount of subdomains within the system and is for most systems much smaller than the transversal discretization. At the boundary of subdomains mode matching methods calculate the coupling of the eigenmodes of one subdomain into the eigenmodes of the next. Complicated structures might require a large number of eigenmodes to correctly model and can therefore be difficult to simulate.
\subsection{Mathematical definitions}
In order to prevent any confusion, we give a quick overview of the mathematical symbols and definitions used in this proceeding. Mathematical objects: matrices are denoted by $\overline{\overline{M}}$, vectors by $\vec{v}$, unit vectors by $\hat{u}$ and scalars by $s$. The identity matrix is denoted by $\overline{\overline{I}}$. For mathematical operations: the conjugate is denoted by $a^*$, the transpose by $\vec{a}^T$ and the conjugate transpose by $\vec{a}^{\dagger}$. We write the dot product as  $\vec{a} \cdot \vec{b}$ and imply that for complex vectors the dot product $\vec{a} \cdot \vec{b}^*$ is performed. Forward propagation is defined by $\exp(-j k_z z)$, with $j = \sqrt{-1}$.
\section{Methods}\label{sec:method}
Here we show how we implemented a modal method with vector plane wave eigenmodes as rigorous solutions to Maxwell's equations in homogeneous isotropic, anisotropic, diattenuating and gyrotropic media. This modal method is based on McLeod\&Wagner (2014)\cite{mcleod2014vector} and is referred to as the $k$-method. It is exact within the Rayleigh-Sommerfeld type I boundary condition approximation that requires the conservation of tangential electric fields at the source plane. The $k$-method includes plane parallel propagation of propagating and evanescent modes, near and far fields, and general Fresnel reflection and transmission coefficients with full polarization effects. Here we will give a summary of the most important ideas and for a more complete overview we refer to their paper\cite{mcleod2014vector}.
\subsection{Vector wave propagation}
The source-free wave equation for a vector electric field propagating in a material with homogeneous real relative dielectric ($\overline{\overline{\epsilon}}'(t)$) and conductivity ($\overline{\overline{\sigma}}(t)$) tensors (assuming non-magnetic media $\overline{\overline{\mu}} = \overline{\overline{I}}$) is: 
\begin{equation}\label{eq:WaveEquation}
\nabla \times \nabla \times \vec{E} + \frac{1}{c^2} \frac{\partial^2}{\partial t^2} [\overline{\overline{\epsilon}}'(t) * \vec{E}] + \mu_0 \frac{\partial}{\partial t}[\overline{\overline{\sigma}}(t) * \vec{E}] = 0.
\end{equation}
By assuming vector plane wave modes of the form: $\vec{E}(t, \vec{r}) = \mathcal{E} \hat{e} e^{j (\omega t - \vec{k} \cdot \vec{r})}$, we can rigorously solve \autoref{eq:WaveEquation} using spatial and temporal Fourier transforms. Every mode is characterized by the wave vector $\vec{k}$ that determines the direction of propagation, the unit-vector polarization $\hat{e}$ and the complex amplitude of the mode $\mathcal{E}$. Applying the Fourier transforms and filling in the vector plane wave solution, reduces \autoref{eq:WaveEquation} to:
\begin{equation}
\left[ \left( \frac{k}{k_0}\right)^2 ( \overline{\overline{I}} - \hat{k}\hat{k}) - \overline{\overline{\epsilon}}   \right] \hat{e} = 0,
\end{equation}
with the dielectric tensor defined as $\overline{\overline{\epsilon}} (\omega) \equiv \overline{\overline{\epsilon}}'(\omega) - j \overline{\overline{\sigma}}(\omega) / \omega \epsilon_0$, $\vec{k} = k \hat{k}$ and $\hat{k}\hat{k}$ the dyadic product of $\hat{k}$. Note that the temporal component is eliminated and from here on we assume one temporal frequency $\omega$ and won't explicit mention it. In order to find the unit-vector polarization $\hat{e}$ of the plane waves, we have to find the unit-vectors that span the null space of the matrix $\overline{\overline{M}} = (k/k_0)^2 ( \overline{\overline{I}} - \hat{k}\hat{k}) - \overline{\overline{\epsilon}} $. To do that, the complete wave vector needs to be determined. By taking the spatial Fourier transform of the field we already determined the transverse components $k_x$ and $k_y$. We can determine $k_z$ by setting the determinant of $\overline{\overline{M}}$ to zero to find the non-trivial solutions. The Booker quartic is the result:
\begin{equation}
a_4 \left( \frac{k_z}{k_0} \right)^4 + a_3 \left( \frac{k_z}{k_0} \right)^3 + a_2 \left( \frac{k_z}{k_0} \right)^2 + a_1 \left( \frac{k_z}{k_0} \right)^1 + a_0 = 0,
\end{equation}
with $a_i$ coefficients\cite{mcleod2014vector} dependent on $\vec{k}_{\perp} \equiv k_x \hat{x} + k_y \hat{y}$, the normal of the surface $\hat{z}$ that points in the z-direction, the dielectric tensor $\overline{\overline{\epsilon}}$ and the length of the wave number in vacuum $k_0 = 2 \pi / \lambda$.\\
\\
\indent The Booker quartic has 4 solutions for every $k_x$, $k_y$ pair, which consist of two forward- and two backward-propagating modes. The coefficients $a_3$ and $a_1$ will be zero for materials with diagonalisable dielectric tensors, resulting in two forward propagating modes and two backward propagating modes that are equal to, but point in the opposite direction w.r.t. the forward propagating modes. If the dielectric tensor is not diagonalisable, the forward and backward propagating modes can not be equal but opposite and the material therefore has not the time reversal property. The uniqueness of the solutions is dependent on the structure of the dielectric tensor. Modes in isotropic materials ( $\overline{\overline{\epsilon}} = \epsilon\overline{\overline{I}}$) propagate in the same direction ($k_{z,1} = k_{z,2}$). Anisotropic materials ($\epsilon_{xx} \neq \epsilon_{yy} \neq \epsilon_{zz}$) on the other hand, will have modes that do not propagate in same direction ($k_{z,1} \neq k_{z,2}$). Solutions with $k_z \in \mathbb{R}$ are plane waves that propagate energy in the z-direction and are called propagating modes. When $k_z \in \mathbb{C}$ and $\Im(k_z) \neq 0$ the modes are evanescent and do not propagate energy.\\  
\\
\indent The null space of $\overline{\overline{M}}$ can now be determined by using any null space finder algorithm. We decided to implement the QR-algorithm\cite{francis1961qr}'\cite{francis1962qr}'\cite{kublanovskaya1962some}, see \autoref{subsec:FindingEigenvectors} for a more complete discussion. Propagating modes in non-attenuating isotropic materials, will have a pair of unit-vector polarizations that form a orthonormal basis. Similarly modes in non-attenuating anisotropic materials will span $\mathbb{R}^3$ but not form a orthonormal basis ($\hat{e}_1 \cdot \hat{e}_2 \neq 1$), as the two modes do not propagate in the same direction. The polarization and the wave vector of all propagating modes in non-attenuating materials must satisfy $\vec{D} \cdot \vec{k} = (\overline{\overline{\epsilon}} \hat{e}) \cdot \vec{k} = 0$. Evanescent modes will span $\mathbb{C}^3$, but do not form a orthonormal basis and will not satisfy $\vec{D} \cdot \vec{k} = 0$\cite{born2013principles}.\\
\\
\indent The propagation direction and polarization of the eigenmode pair at $k_x, k_y$ have now been determined. They can be used to expand the input electromagnetic field $\vec{E}(x,y)$ on the eigenmodes:
\begin{align}
\mathcal{E}_1(k_x, k_y) &= \mathcal{F}_{x,y} \left\{ \vec{E}(x,y)  \right\} \cdot \vec{p}_1, \\
				     &= \mathcal{F}_{x,y} \left\{ \vec{E}(x,y)  \right\} \cdot \frac{\hat{e}_2 \times \hat{z}}{(\hat{e}_1 \times \hat{e}_2) \cdot \hat{z}}, \nonumber
\end{align}
\begin{align}
\mathcal{E}_2(k_x, k_y) &= \mathcal{F}_{x,y} \left\{ \vec{E}(x,y)  \right\} \cdot \vec{p}_2, \\ 
				     &= \mathcal{F}_{x,y} \left\{ \vec{E}(x,y)  \right\} \cdot \frac{\hat{z} \times \hat{e}_1}{(\hat{e}_1 \times \hat{e}_2) \cdot \hat{z}}, \nonumber
\end{align}
with $\hat{e}_i$ the unit-vector polarization corresponding to the eigenmode $i$ and $\vec{p}_i$ a projection vector\cite{mcleod2014vector} that satisfies the Rayleigh-Sommerfeld type I boundary conditions. The field coefficients of the two propagating modes can be put into a vector:
\begin{equation}
\vec{x} = 
\begin{pmatrix}
	\mathcal{E}_1 \\
	\mathcal{E}_2 \\
\end{pmatrix}.
\end{equation}
The propagation of the modes can be described by a matrix multiplication with the 2x2 propagation matrix:
\begin{equation}\label{eq:Propagation_Tensor}
\overline{\overline{P}}(z, k_{z,1}, k_{z,2}) = 
\begin{pmatrix}
	e^{-j k_{z,1}z} & 0 \\
	0 & e^{-j k_{z,2}z}
\end{pmatrix}.
\end{equation}
The field in the eigenmodes propagated after a distance $z$ is then given by:
\begin{equation}
\vec{x}' = \overline{\overline{P}}(z, k_{z,1}, k_{z,2}) \vec{x}.
\end{equation}
\subsection{General Fresnel coefficients}\label{subsec:General_Fresnel_Equations}
Plane waves will only couple into other plane waves with identical transverse spatial frequencies when they encounter a boundary. This is due to the phase matching requirement\cite{born2013principles} set by the boundary conditions. Using the polarization unit-vectors and the wave vectors of the eigenmodes in combination the boundary conditions, which state that the tangential electric and magnetic field must be continuous. We find that the equations that must be solved to determine the reflected and transmitted fields are:
\begin{align}
\hat{e}_{n,\perp}^i + \rho_{1n} \hat{e}_{1,\perp}^r + \rho_{2n} \hat{e}_{2,\perp}^r &= \tau_{1n} \hat{e}_{1,\perp}^t + \tau_{2n} \hat{e}_{2,\perp}^t, \\
\vec{h}_{n,\perp}^i + \rho_{1n}  \vec{h}_{1,\perp}^r +\rho_{2n} \vec{h}_{2,\perp}^r &=\tau_{1n} \vec{h}_{1,\perp}^t + \tau_{2n} \vec{h}_{2,\perp}^t, \nonumber
\end{align}
with the magnetic polarization given by $\vec{h}_m \equiv \hat{e}_m \times \vec{k}_m$. Note that the incoming mode $\hat{e}_n^i$ can couple in both reflected and transmitted modes $\hat{e}^{r/t}_1$ and $\hat{e}^{r/t}_2$. Reflection and transmission will therefore induce modal crosstalk. These equations can be solved and the analytical solutions that we have implemented are given by Macleod\&Wagner\cite{mcleod2014vector}. The solutions are completely general for any boundary between two materials described by dielectric tensors, both for the propagating and evanescent regimes for which continuity of tangential fields is correct. This might not be the case at the boundary of optically active media, it has been proposed that the normal components of the magnetic field should then be conserved instead of the tangential. The general Fresnel coefficients given by reflection and transmission $2\times2$ matrices are:
\begin{equation}\label{eq:reflection_transmission_tensors}
\overline{\overline{\tau}} = 
\begin{pmatrix}
	\tau_{11} & \tau_{12} \\
	\tau_{21} & \tau_{22} 
\end{pmatrix}
, \ \overline{\overline{\rho}} = 
\begin{pmatrix}
	\rho_{11} & \rho_{12} \\
	\rho_{21} & \rho_{22}.
\end{pmatrix}
\end{equation}
Therefore the forward propagating modes couple into the new forward and backward propagating modes by:
\begin{equation}\label{eq:Reflection_and_Transmission}
\vec{x}_t = \overline{\overline{\tau}} \vec{x}_i, \ \ \vec{x}_r = \overline{\overline{\rho}} \vec{x}_i.
\end{equation}
\subsection{Interference}
\begin{figure}
\begin{center}
\begin{tabular}{c}
\includegraphics[height=4cm]{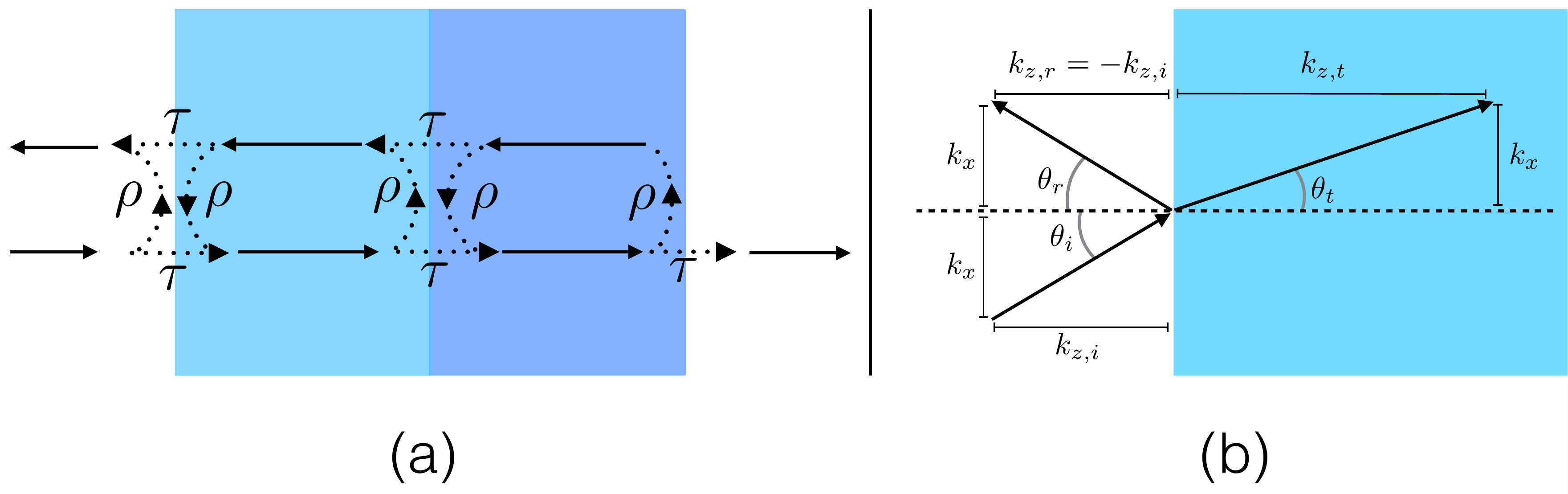}
\end{tabular}
\end{center}
\caption 
{ \label{fig:Interference_and_Snells_Law}
(a) A schematic overview of interference effects in a stack of two thin films which are illuminated from the left. The schematic shows the complexity of the possible paths that can be taken within three boundaries. (b) A schematic of the wave vector components when refracting into a new material. The phase matching requirement results in the conservation of the transverse components of the wave vector.} 
\end{figure} 
Optics consisting of stacks of thin films are important within many optical systems. For example anti- or high-reflection coatings improve the transmission or reflectivity of surfaces. Interference effects within these optics occur when part of the electromagnetic waves propagating through the optic undergoes multiple internal reflections (\autoref{fig:Interference_and_Snells_Law} (a)). Through interference with the incoming waves, it enhances the transmission or reflectivity of the entire optical element. Within wave plates the interference is also polarization dependent and is, for example, limiting the sensitivity of spectropolarimetry\cite{semel2003spectropolarimetry}.\\ 
\indent In the above subsections we identified $2\times2$ matrices that describe the propagation and reflection/refraction effects on the two modes. This are the necessary ingredients to implement the interference effects.\\ 
\indent We have implemented the recursive S-matrix algorithm\cite{lavrinenko2015numerical} with the assumption that we have incident light from one side, see \autoref{fig:Interference_and_Snells_Law} (a). For every layer, the algorithm calculates one round trip of a wave within the layer. Summing roundtrips to infinity results into a converging geometrical series, which can be reduced to a simple form when the absolute values of the eigenvalues are below unity. This is recursively done  for all layers, adding them together layer by layer, until two $2\times2$ matrices $\overline{\overline{t}}$ and $\overline{\overline{r}}$ describe the reflection and transmission properties of the entire stack. The complex amplitude of the reflected and transmitted fields is calculated as in \autoref{eq:Reflection_and_Transmission}.\\
\indent Li (1996)\cite{li1996formulation} has shown that our current implementation of the S-matrix algorithm might not stable for evanescent waves. To ensure stability, an algorithm that is numerical stable needs to be implemented\cite{li1996formulation}. Our implementation, combining the S-matrix algorithm with the $k$-method, is similar to Berreman calculus \cite{berreman1972optics}'\cite{mccall2014birefringent}'\cite{weenink2012application}.
\subsection{Electric field reconstruction}\label{subsec:Efield_Reconstruction}
Using the above introduced methods, we can now construct a system that consists of materials linked together by parallel boundaries. All operations like propagation, reflection and refraction are described by $2\times2$ matrices. Therefore, the entire system can now be be put together by matrix multiplication. A system with $N$ operations is given by:
\begin{equation}
\overline{\overline{J}}_{sys} (k_x, k_y) = \overline{\overline{J}}_N (k_x, k_y) \overline{\overline{J}}_{N-1}(k_x, k_y) \dots \overline{\overline{J}}_2 (k_x, k_y)\overline{\overline{J}}_1 (k_x, k_y),
\end{equation}
with $\overline{\overline{J}}_1 (k_x, k_y)$ the first operation in the system, $\overline{\overline{J}}_2 (k_x, k_y)$ the second, etc. The resulting matrix $\overline{\overline{J}}_{sys}(k_x, k_y)$ describes the behaviour of the system as a function of spatial frequency. When $\overline{\overline{J}}_{sys}(k_x, k_y)$ is calculated at the focal plane, it is is similar to the Jones pupil\cite{breckinridge2015polarization}'\cite{ruoff2009orientation} conjugated to that focal plane. The behaviour of the system in the spatial domain $\overline{\overline{J}}_{sys}$($x$,$y$) is obtained by taking the inverse Fourier transform of the separate matrix elements. When done at the focal plane, it is equivalent to the amplitude response matrix (ARM), which is defined by McQuire\&Chipman (1990)\cite{mcguire1990diffraction}. From now we will adopt the nomenclature of Jones pupil and ARM.\\
\begin{equation}
\overline{\overline{J}}_{ARM} = \mathcal{F}^{-1}_{k_x, k_y}\left\{ \overline{\overline{J}}_{pupil} \right\} = 
\begin{pmatrix}
	\mathcal{F}^{-1}_{k_x, k_y}\left\{J_{11}\right\} & \mathcal{F}^{-1}_{k_x, k_y}\left\{J_{12}\right\} \\
	\mathcal{F}^{-1}_{k_x, k_y}\left\{J_{21}\right\} & \mathcal{F}^{-1}_{k_x, k_y}\left\{J_{22} \right\}
\end{pmatrix}
\end{equation} 
High resolution ARM's can be created by performing the partial discrete Fourier transform \cite{soummer2007fast}. The complex amplitudes of the eigenmodes after propagation through the system are calculated as:
\begin{equation}
\begin{pmatrix}
	\mathcal{E}_1^o \\
	\mathcal{E}_2^o \\
\end{pmatrix}
= \overline{\overline{J}}_{pupil}
\begin{pmatrix}
	\mathcal{E}_1^i \\
	\mathcal{E}_2^i \\
\end{pmatrix}.
\end{equation}
From these complex amplitudes the electric field is then reconstructed:
\begin{equation}
\vec{E}(\vec{r}) = \mathcal{F}^{-1}_{k_x, k_y} \left\{ \sum_{i = 1, 2} \hat{e}_i \mathcal{E}^o_i \right\},
\end{equation}
with $\hat{e}_i$ the unit-vector polarizations of the material in which the electric field is reconstructed.
\subsection{Unpolarized light}
The Jones pupil and ARM act on light that is completely polarized and are therefore deterministic processes. However, unpolarized light is a statistical phenomenon and therefore analyzation of system effects on unpolarized requires a slightly different approach. We implemented two different methods to handle unpolarized light: coherency matrices and Mueller matrices.
\subsubsection{Coherency matrices}
The coherency matrix\cite{wolf2007introduction} is an equal-time correlation matrix between the two modes and is given by:
\begin{equation}
\overline{\overline{\Phi}} = 
\begin{pmatrix}
	 \langle \mathcal{E}_1 \mathcal{E}_1^* \rangle & \langle \mathcal{E}_1 \mathcal{E}_2^* \rangle \\
  	\langle \mathcal{E}_2 \mathcal{E}_1^* \rangle & \langle \mathcal{E}_2 \mathcal{E}_2^* \rangle
\end{pmatrix}. 
\end{equation}
Conveniently, Pauli matrices can be used as a basis for coherency matrices:
\begin{equation}
\overline{\overline{\Phi}} = \frac{1}{2} \sum_{i=0}^3 S_i \overline{\overline{\sigma_i}},
\end{equation}
with the Pauli matrices given by:
\begin{equation}
\overline{\overline{\sigma}}_0 = 
\begin{pmatrix}
	1 & 0 \\
	0 & 1
\end{pmatrix}, \ 
\overline{\overline{\sigma}}_1 = 
\begin{pmatrix}
	1 & 0 \\
	0 & -1
\end{pmatrix}, \
\overline{\overline{\sigma}}_2 = 
\begin{pmatrix}
	0 & 1 \\
	1 & 0
\end{pmatrix}, \
\overline{\overline{\sigma}}_3 = 
\begin{pmatrix}
	0 & -i \\
	i & 0
\end{pmatrix}.
\end{equation}
The amplitudes $S_i$ that are multiplied with the basis to construct the coherency matrix are the Stokes parameters: $S_0 = I$, $S_1 = Q$, $S_2 = U$, $S_3 = V$. Completely unpolarized light is given by $\overline{\overline{\sigma}}_0/2$, completely $+Q$ polarized light by $(\overline{\overline{\sigma}}_0 + \overline{\overline{\sigma}}_1)/2$, etc. A system described by $\overline{\overline{J}}_{ARM}$ has a relation between $\overline{\overline{\Phi}}_{in}$ and $\overline{\overline{\Phi}}_{out}$:
\begin{equation}
\overline{\overline{\Phi}}_{out} = \overline{\overline{J}}_{ARM} \overline{\overline{\Phi}}_{in} \overline{\overline{J}}_{ARM}^{\dagger}.
\end{equation}
\subsubsection{Mueller matrices}
Another way to also analyse the performance of the system with unpolarized light is by converting $2\times2$ matrices to the Mueller matrix formalism. This conversion is given by\cite{bass2009handbook}:
\begin{equation}
\overline{\overline{M}} = \overline{\overline{U}} (\overline{\overline{J}} \otimes   \overline{\overline{J}} ^*)  \overline{\overline{U}}^{-1},
\end{equation}
with $\overline{\overline{U}}$:
\begin{equation}
\overline{\overline{U}}  = \frac{1}{\sqrt{2}}
\begin{pmatrix}
	1 & 0 & 0 & 1 \\
  	1 & 0 & 0 & -1 \\
  	0  & 1 & 1 & 0  \\
  	0 & i & -i & 0 
\end{pmatrix}. 
\end{equation}
Converting the ARM to a Mueller matrix gives the Point Spread Matrix (PSM)\cite{mcguire1990diffraction}. Analogous we can also calculate the Mueller pupil by converting the Jones pupil. There is no Fourier relation between the Mueller pupil and the PSM.\\
\indent Note that in the Mueller matrix formalism absolute phase information is lost. In the coherency matrix formalism the system is still described by Jones matrices and therefore the phase information is retained.
\section{Numerical implementation}\label{sec:Numerical_Implementation}
In this section we describe how we have implemented the methods described in \autoref{sec:method} in Python3.  We will compare how the implementation of the $k$-method compares to existing analytical expressions.
\subsection{Solving the Booker quartic}\label{subsec:solvingBooker}
\begin{figure}
\begin{center}
\begin{tabular}{c}
\includegraphics[height=8cm]{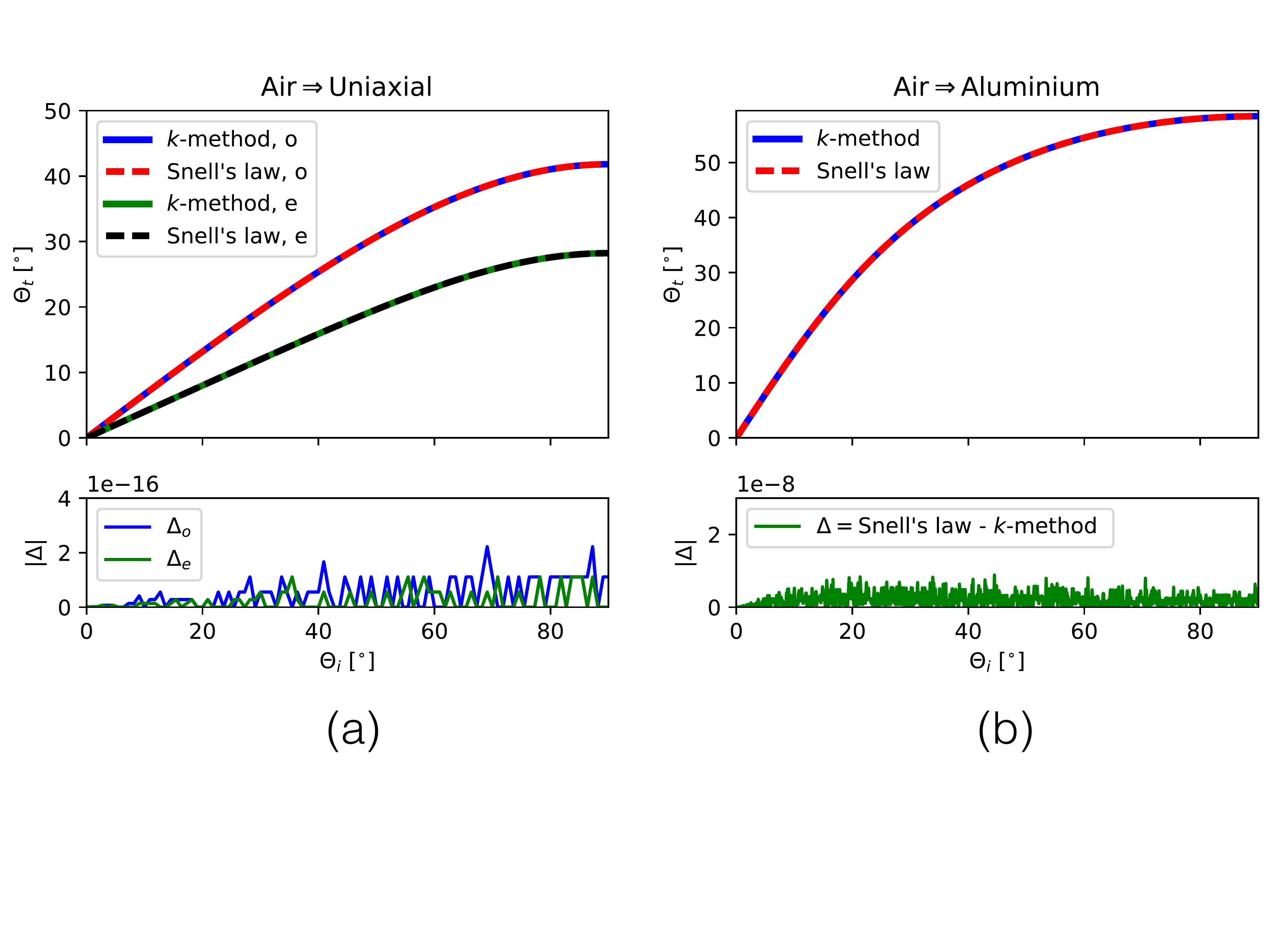}  
\end{tabular}
\end{center}
\caption 
{ \label{fig:Snells_Law}
A comparison between the $k$-method and analytical expression for Snell's law: (a) different propagation directions of the ordinary (o) and extraordinary (e) waves refracting into an anisotropic material ($n_e = 2.5$, $n_o = 1.5$). The numerical accuracy is on the order of $\sim 10^{-16}$. (b) refraction from air into aluminium ($n = 0.62569 - 5.3205j$), with a numerical accuracy of $\sim 10^{-8}$.} 
\end{figure}

\begin{figure}
\begin{center}
\begin{tabular}{c}
\includegraphics[height=7cm]{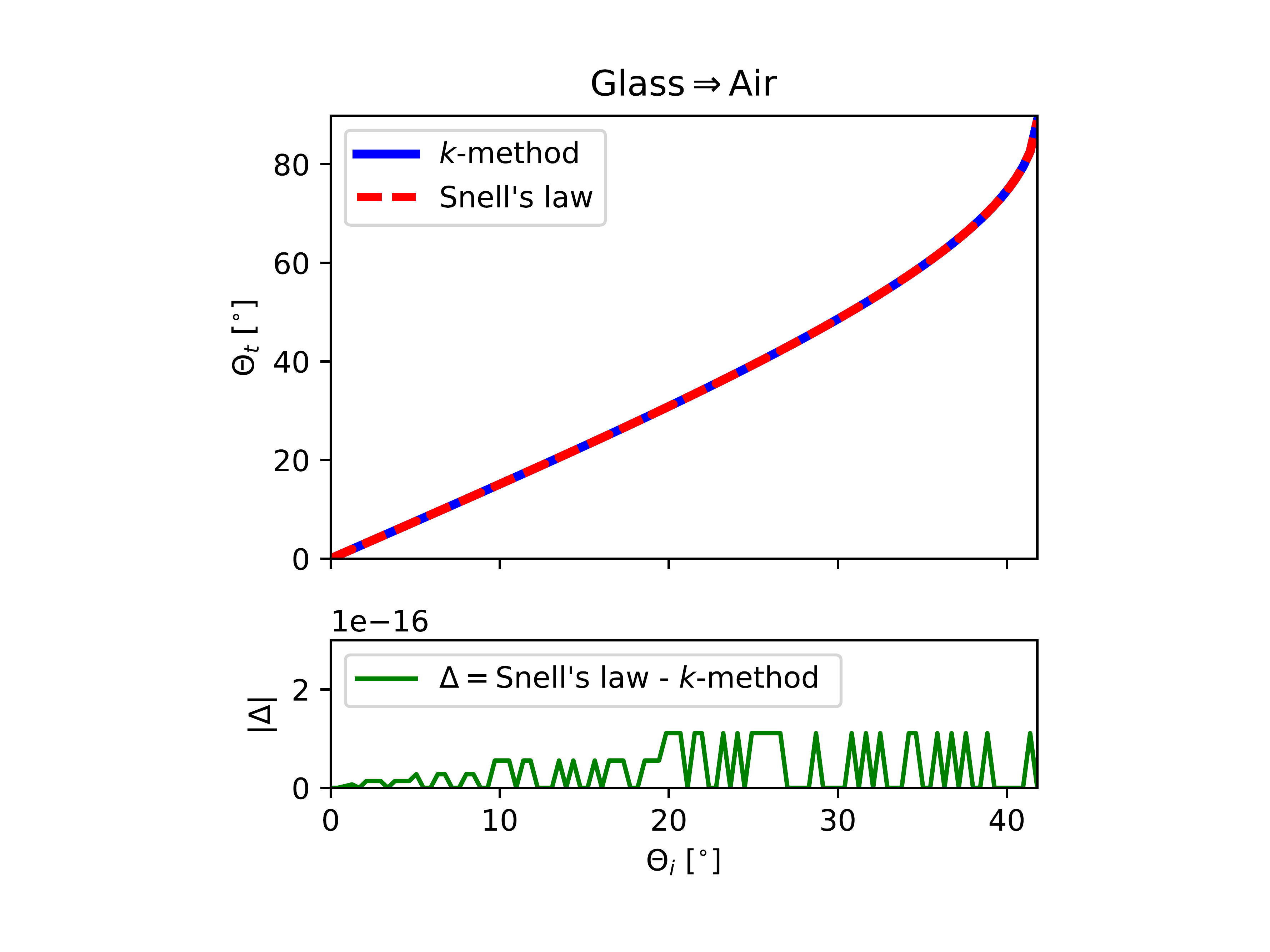}  
\end{tabular}
\end{center}
\caption 
{ \label{fig:Snells_Law_2}
A comparison between the $k$-method and analytical expression for Snell's law: total internal reflection at the boundary between glass ($n = 1.5$) and air ($n=1$). The numerical precision is on the order of $\sim 10^{-16}$.} 
\end{figure}

General solutions to quartic equations are known since the 16$^{th}$ century. Famous mathematicians such as Lodovico Ferrari, Ren\'{e} Descartes and Leonhard Euler all came up with their own solution. Therefore solving the Booker quartic is a matter of selecting one of many methods. As the solutions are analytical, we expect numerical errors on the level of numerical precision $\sim 10^{-15}-10^{-16}$.\\ 
\indent We can verify our solutions of the Booker's quartic by applying phase matching at a boundary, i.e. conservation of the transverse wave vector:
\begin{equation}
k_{x,i} = k_{x,t} = k_x,
\end{equation}
to determine the angle of refraction as a function of angle of incidence. See \autoref{fig:Interference_and_Snells_Law} (b) for a schematic. Then comparing the results with the analytical solution derived from phase matching: Snell's law of refraction.\\
\begin{equation}\label{eq:Snells_Law}
n_1 \sin(\theta_i) = n_2 \sin(\theta_t)
\end{equation}
\indent In \autoref{fig:Snells_Law} and \autoref{fig:Snells_Law_2} the $k$-method is compared to Snell's law for various sets of materials. The $k$-method and Snell's law agree until numerical precision for non-absorbing materials, i.e. the angle of total internal reflection is reproduced (\autoref{fig:Snells_Law_2}) and the splitting of two modes when refracting into an birefringent material is reproduced as well (\autoref{fig:Snells_Law} (a)). The algorithm has a lower numerical accuracy for absorbing materials, see \autoref{fig:Snells_Law} (b). Putting the solutions for absorbing materials back into the Booker quartic results in residuals on a $\sim10^{-8}$ level. Therefore we suspect that this lower precision is due numerical effects in the algorithm solving the quartic equation. We were unable to pinpoint the exact source of the lower precision.\\
\subsection{Determining the eigenvectors}\label{subsec:FindingEigenvectors}
Finding the basis of the null space of the matrix $\overline{\overline{M}}$, given by: 
\begin{equation}
\overline{\overline{M}} = \left( \frac{k}{k_0} \right)^2 ( \overline{\overline{I}} - \hat{k}\hat{k}) - \overline{\overline{\epsilon}} 
\end{equation}
results in the unit-vector polarizations that form the basis of the electric field modes in the material for a given  transverse spatial frequency pair ($k_x$, $k_y$). The solutions of the Booker quartic are analytic and can be simultaneous computed for a grid of transverse spatial frequencies. In Python this greatly reduces the computation time for large grids. Therefore a similar algorithm for the unit-vector polarizations is preferred.\\ 
\indent For isotropic materials we have written out the analytical expressions for two different bases (s-, p- and $E_x$, $E_y$ polarization modes). The $E_x$, $E_y$ polarization modes are given by:
\begin{equation}\label{eq:ExEyModes}
\hat{e}_1 = C_1 
\begin{pmatrix}
	0 \\
	k_z \\
	-k_y
\end{pmatrix}, \ \ 
\hat{e}_2 = C_2 
\begin{pmatrix}
	k_y^2 + k_z^2 \\
	-k_x k_y \\
	-k_x k_z
\end{pmatrix}.
\end{equation}
The s- and p-polarization modes are given by:
\begin{equation}
\hat{e}_1 = C_1
\begin{pmatrix}
	1 \\
	-k_x/k_y \\
	0
\end{pmatrix}, \ \ 
\hat{e}_2 = C_2
\begin{pmatrix}
	k_x k_z / k_y \\
	k_z \\
	- k_y - k_x^2/k_y	
\end{pmatrix},
\end{equation}
with normalization constants $C_1$ and $C_2$ such that $|\hat{e}_1|^2 = |\hat{e}_2|^2= 1$. These solutions form an orthonormal basis for the propagating modes ($k_z \in \mathbb{R}$, $\hat{e}_i\cdot\hat{e}_j = \delta_{ij}$) and are perpendicular to $\vec{k}$ ($\hat{e}_i \cdot \vec{k}_i = 0$). For evanescent modes ($k_z \in \mathbb{C}$) these vectors form a linear independent basis ($\hat{e}_1 \cdot \hat{e}_2 \neq 1$) and $\hat{e}_i \cdot \vec{k}_i \neq 0$\cite{born2013principles}. The unit-vector polarizations of non-absorbing isotropic materials satisfy $\overline{\overline{M}}_i \cdot \hat{e}_i = 0$ to numerical precision $\sim10^{-15}$. For absorbing isotropic materials we are limited to $\sim10^{-8}$, which is due to the precision in the $k_z$ solutions (\autoref{subsec:solvingBooker}).\\
\begin{figure}
\begin{center}
\begin{tabular}{c}
\includegraphics[height=7cm]{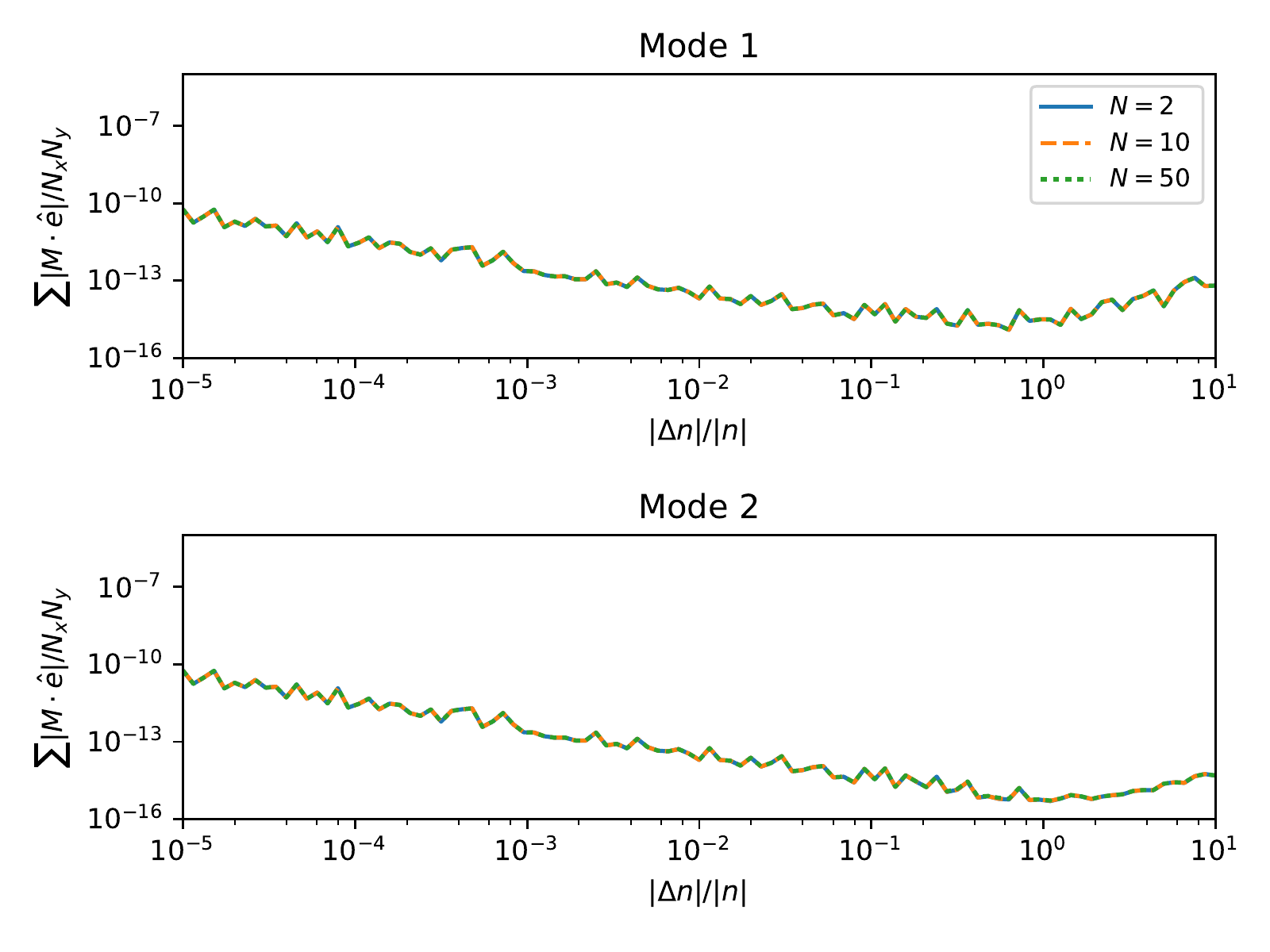}
\end{tabular}
\end{center}
\caption 
{ \label{fig:Stability_QR}
The numerical precision of the QR algorithm to find the null space of $\overline{\overline{M}}$ for the two modes. The merit is an average of the three components of the unit-vector polarization and the spatial frequencies sampled.} 
\end{figure}
\indent The basis of unit-vector polarizations for anisotropic materials is determined by the QR algorithm\cite{francis1961qr}'\cite{francis1962qr}'\cite{kublanovskaya1962some}. The QR decomposition is performed by Givens rotations\cite{bindel2002computing} with the rotation matrices $\in$ SU(2) combined with $r = \sqrt{|c|^2 + |s|^2}$:
\begin{equation}
\begin{pmatrix}
	c & s \\
	-s^* & c 
\end{pmatrix} 
\begin{pmatrix}
	a \\
	b
\end{pmatrix} = 
\begin{pmatrix}
	r \\
	0
\end{pmatrix}
\end{equation}
These matrices ensure that for complex elements in $\overline{\overline{M}}$ (evanescent waves), the rotations are performed as expected. In \autoref{fig:Stability_QR} the stability of the QR-algorithm is plotted with changing birefringence of a uniaxial material. The algorithm performs optimally around a $|\Delta n|/ |n|$ of $1$ , achieving a precision of $\sim10^{-15}$. Different colours depict different numbers of QR decompositions performed and these do not show any improvement. This suggests that this is the limit in accuracy of the current implementation of the QR algorithm. There are two structures that dominate the curve. The first structure is the decreasing numerical error with increasing $|\Delta n|/ |n|$ for $|\Delta n|/ |n| < 1$. This is likely due to the bare QR algorithm not globally converging when the ratio of the eigenvalues becomes close to one, which is the case for small birefringence. Implementing shift strategies\cite{wang2002convergence} might result in global convergence, but are not included in this work. The second structure is the decrease in performance for $|\Delta n|/ |n| > 1$. This is due to an unexpected decrease in numerical accuracy of our implementation of the matrix adjoint operator necessary for the coefficients of the Booker quartic. This is likely due to the matrix inverse and trace operations affected by large differences in the eigenvalues. \\
\subsection{Validation of the Generalised Fresnel coefficients}\label{subsec:Fresnel_Coef}
\begin{figure}
\begin{center}
\begin{tabular}{c}
\includegraphics[height=12cm]{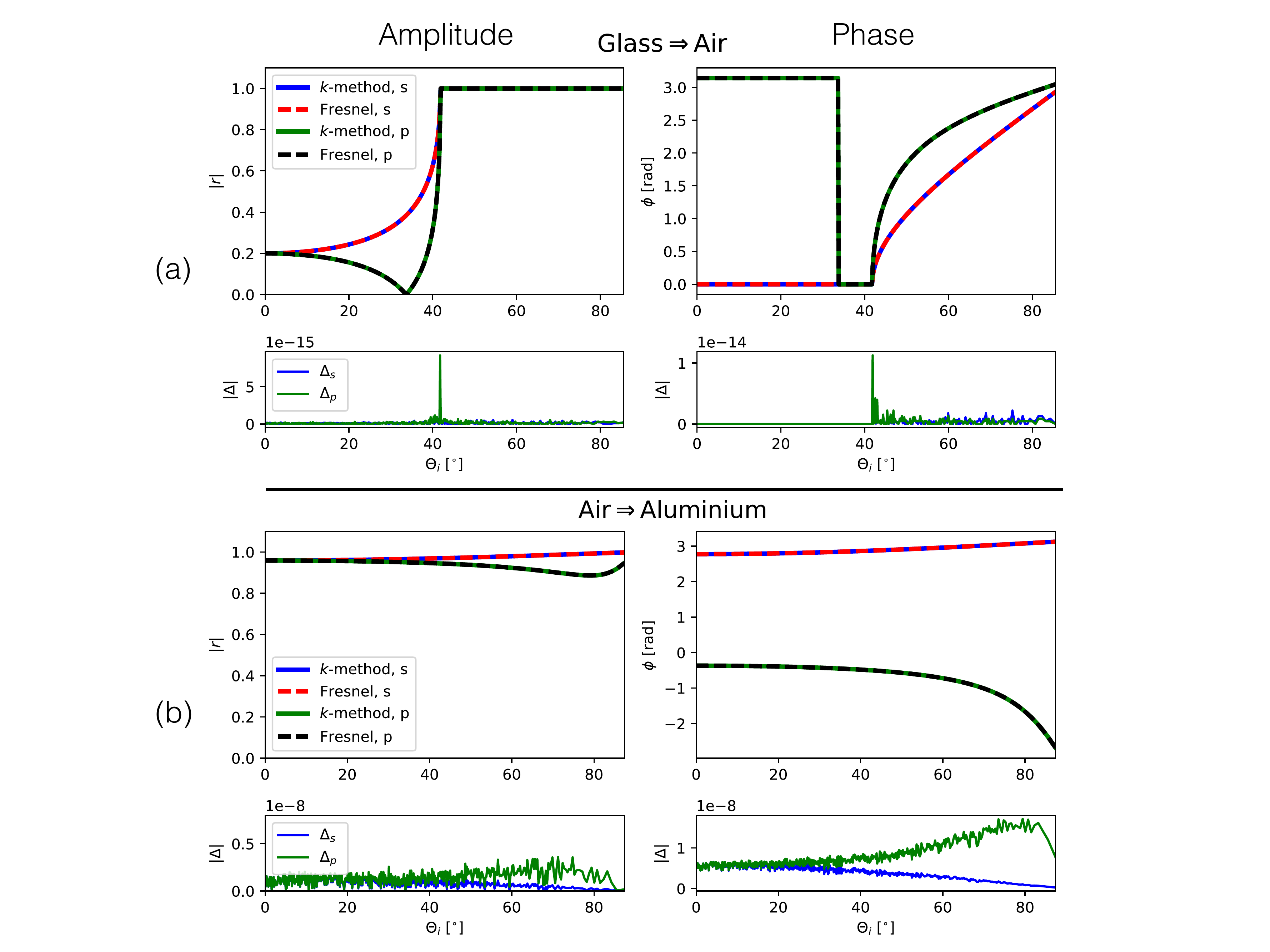}  
\end{tabular}
\end{center}
\caption 
{ \label{fig:Fresnel_Coefficients_Glass_Air}
A comparison of the amplitude and phase effects between Fresnel equations and $k$-method for (a) glass to air and (b) air to aluminium. Both methods agree to a $\sim 10^{-15}-10^{-16}$ level for the boundary between glass and air and on a $\sim10^{-8}$ level for air to aluminium.} 
\end{figure}

In \autoref{fig:Fresnel_Coefficients_Glass_Air} the comparison between Fresnel coefficients of the $k$-method and Fresnel's equations is calculated for various materials. \autoref{fig:Fresnel_Coefficients_Glass_Air} (a) shows the boundary between glass and air with an error on the order of numerical precision $\sim10^{-16}$. We find that the Brewster angle is correctly reproduced and the phase effects of total internal reflection that give rise to the spatial Goos-H\"{a}nchen shift\cite{aiello2009duality}'\cite{lvovsky2013fresnel} are included as well. \\
\indent The Fresnel equations for the boundary between a non-absorbing and absorbing material differ on the order of $\sim10^{-8}$, which is again due to the precision in $k_z$ solutions (\autoref{subsec:solvingBooker}). The amplitude and phase effects resulting in the spatial and angular Goos-H\"{a}nchen shifts\cite{aiello2009duality}'\cite{merano2007observation} are included.\\
\\
\indent Beside an accurate reproduction of the Fresnel equations discussed above, the energy needs to be conserved as well for all spatial frequencies. In \autoref{fig:Energy_Conservation} (a) the reflectance and transmission of a boundary between air and an anisotropic material ($n_e = 2$, $n_o = 1.5$) is shown for all propagating modes. The energy is conserved to numerical precision ($\sim10^{-16}$) for both modes.\\  
\indent For the boundary between air and an absorbing materials (not shown) we find that the energy is conserved to numerical precision $\sim10^{-16}$ as well. This is expected as the analytical solution discussed in \autoref{subsec:General_Fresnel_Equations} will find transmission and reflection coefficients that conserve energy irrespective of numerical accuracy in $\hat{e}$ and $k_z$.
\subsection{Validation of the interference}\label{subsec:ValidationInterference}
\begin{figure}
\begin{center}
\begin{tabular}{c}
\includegraphics[height=10cm]{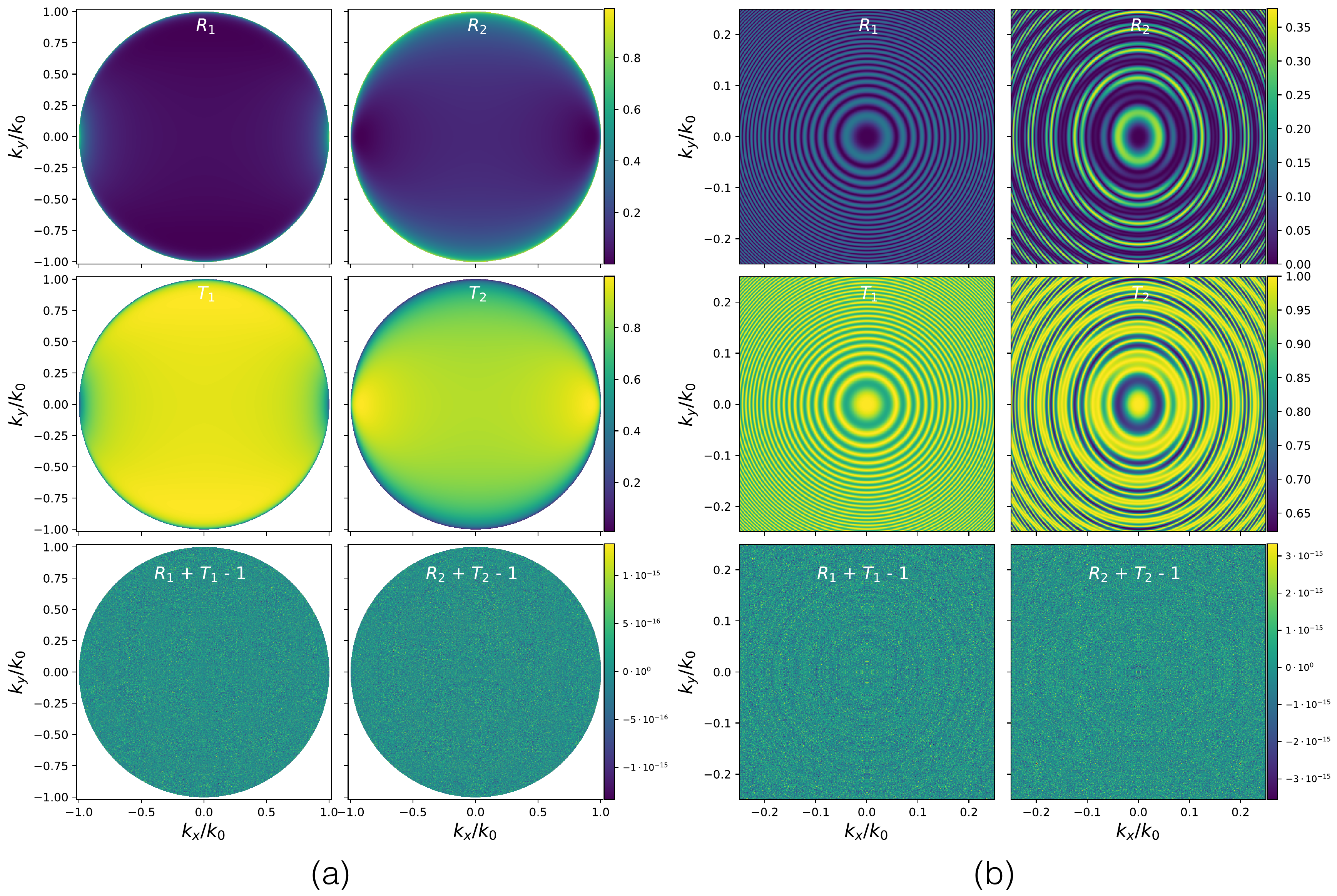}  
\end{tabular}
\end{center}
\caption
{\label{fig:Energy_Conservation}
(a) The transmission $T$ (first row) and reflectance $R$ (second row) of the two modes when reflecting/transmitting of/in a boundary between an isotropic material ($n=1$) and an anisotropic material ($n_e = 2$, $n_o = 1.5$). The third row shows the sum of $R$ and $T$ minus 1, which shows that the results deviate from energy conservation on a $10^{-16}$ level. (b) The transmittance and reflectance are shown for the interference effects within a stack of plates. The stack consists of glass ($n = 1.5$), the anisotropic material from (a) and another plate of glass. The glass plate thicknesses are $50$ $\mu$m, the anisotropic plate is $125$ $\mu$ and  $\lambda = 500$ nm. The fringe structure due to internal reflections is clearly visible and the energy is conserved to a $\sim10^{-16}$ level.} 
\end{figure}

Using energy conservation the interference is validated. In \autoref{fig:Energy_Conservation} (b) the total reflectance and transmittance of a stack wave plate shown. The stack consists of a $50$ $\mu$m thick glass plate ($n = 1.5$) ,  a $125$ $\mu$m thick anisotropic plate (similar as in \autoref{subsec:Fresnel_Coef}) and another plate of glass with the same thickness. The first and second row show respectively the reflectance and transmittance of the two modes. In the third row the deviation from energy conservation is plotted and it shows that energy is conserved to numerical precision $\sim 10^{-16}$.\\
\indent Qualitatively the structure of the fringes in \autoref{fig:Energy_Conservation} (b) agrees with fringe structures seen in Fabry-P\'{e}rot etalons\cite{hecht2002optics}. The figure shows constructive and destructive interference fringes. 
Mode 1 is the ordinary mode and has the classical circular fringe structure. It only experiences internal reflections at the outer boundaries as the ordinary refractive index is equal to the refractive index of glass and therefore no reflections occur at the inner boundaries. Mode 2 is the extraordinary mode and has a more complex elliptical fringe structure. This is due to more complex internal reflections and the changing refractive index with spatial frequency.\\
\section{Example: Mirror reflection}
\begin{figure}
\begin{center}
\begin{tabular}{c}
\includegraphics[height=5cm]{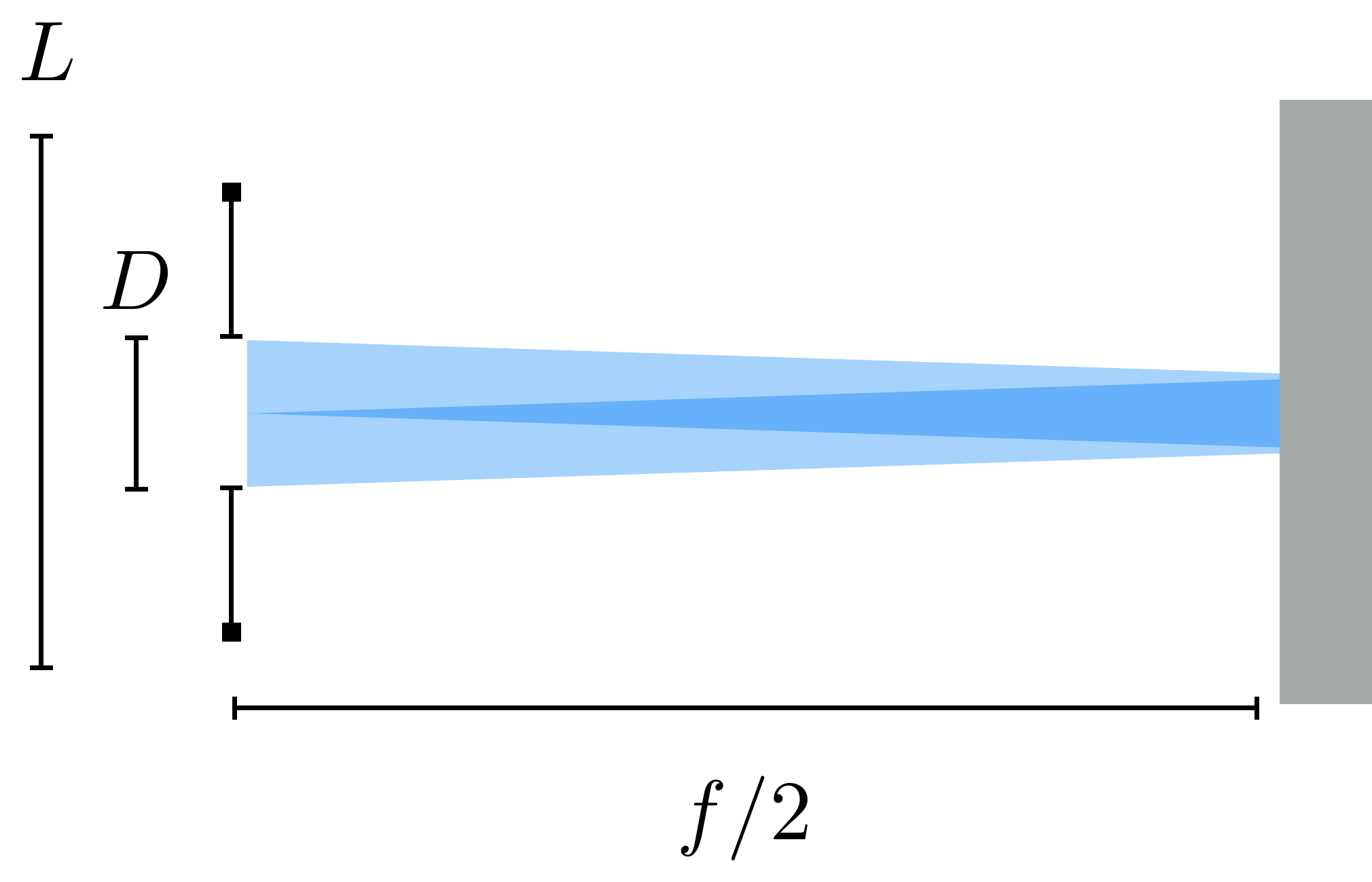}
\end{tabular}
\end{center}
\caption 
{ \label{fig:Simulation_Setup}
The setup of the simulation. The parameters are: the size of the computational domain $L = 400$ $\mu$m, the sampling along one axis $N = 1001$, the diameter of the pupil $D = 25$ $\mu$m, the focal length $f = 100$ $\mu$m.  The f-number of the converging beam is $f_{\#} = 2$. The system is simulated at a wavelength of $500$ nm.  The material of the mirror is aluminium ($n = 0.62569 - 5.3205j$ \cite{mcpeak2015plasmonic}). The system is evaluated in the focal plane.} 
\end{figure}

To demonstrate the current capabilities of the code, we simulate a converging beam reflected by a flat aluminium mirror at normal incidence. The properties of the system are then analyzed in the focal plane. A schematic of the setup is shown in \autoref{fig:Simulation_Setup}.\\
\subsection{Results}
\begin{figure}
\begin{center}
\begin{tabular}{c}
\includegraphics[height=15cm]{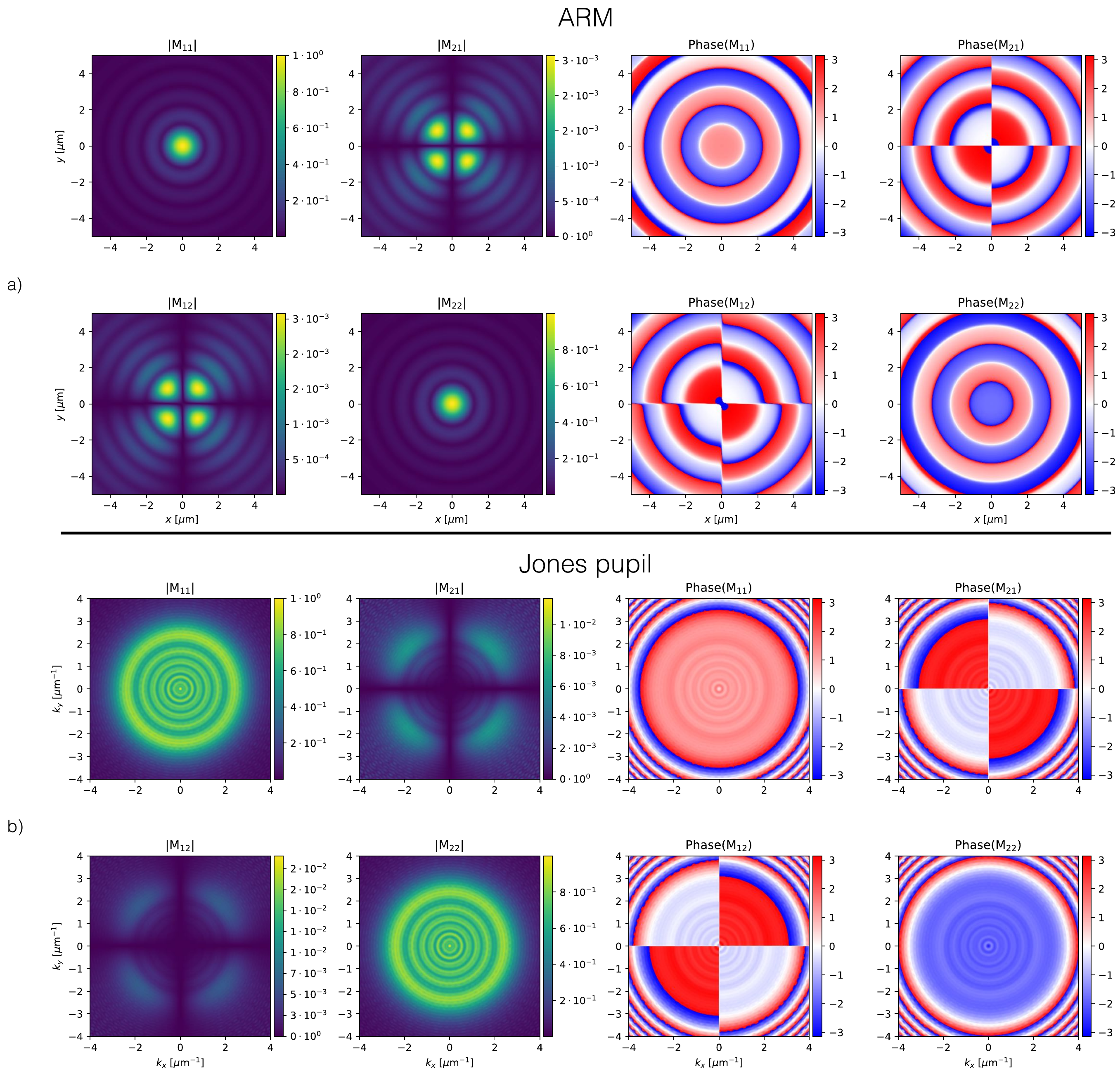}
\end{tabular}
\end{center}
\caption 
{ \label{fig:Jones}
The amplitude and phase effects of the (a) monochromatic ($\lambda = 500$ nm) ARM and (b) Jones pupil of the system. Both figures are normalized to 1.} 
\end{figure}

\begin{figure}
\begin{center}
\begin{tabular}{c}
\includegraphics[height=11cm]{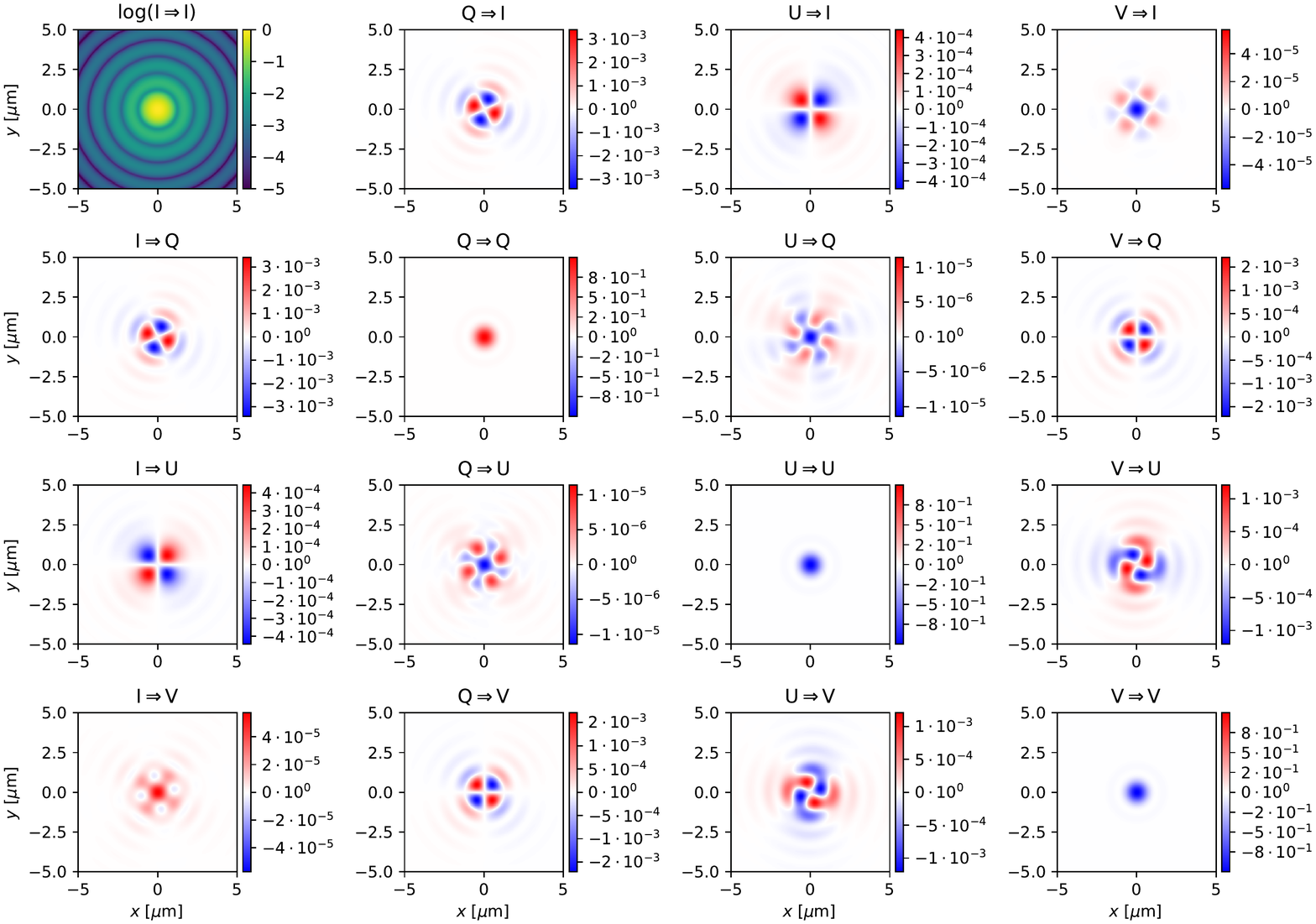}
\end{tabular}
\end{center}
\caption 
{ \label{fig:PSM}
The monochromatic ($\lambda = 500$ nm) PSM of the system. The PSM is normalized to $I \Rightarrow I$.} 
\end{figure}

\begin{figure}
\begin{center}
\begin{tabular}{c}
\includegraphics[height=11cm]{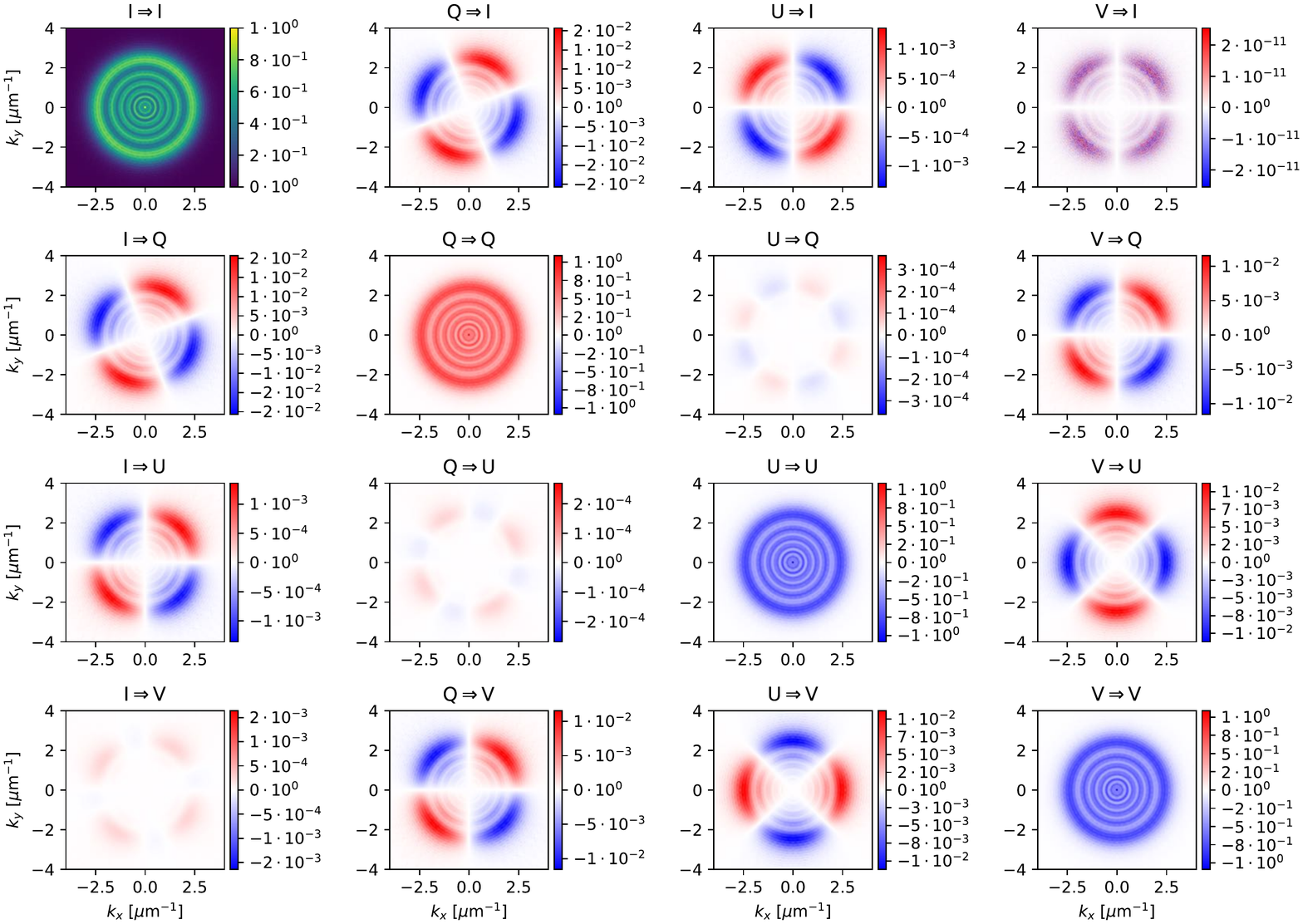}
\end{tabular}
\end{center}
\caption 
{ \label{fig:Mueller_Pupil}
The monochromatic ($\lambda = 500$ nm) Mueller pupil of the system. The Mueller pupil normalized to $I \Rightarrow I$. Note that there is no Fourier relation between the PSM and the Mueller pupil.} 
\end{figure}

\autoref{fig:Jones} shows the ARM and the Jones pupil of the system. The diagonal elements in the ARM are the Airy disk patterns, which are expected from diffraction effects of a circular aperture. The off-diagonal elements have a more complicated structure with much lower amplitudes. These structures are due to modal cross-talk when the beam reflects from the surface. We see in the Jones pupil that the cross-talk mainly originates from reflections at relatively high spatial frequencies.\\
\indent \autoref{fig:PSM} and \autoref{fig:Mueller_Pupil} show the expansion of the ARM and Jones pupil into the PSM and Mueller pupil. We see in the PSM (\autoref{fig:PSM}) that the mirror has structure in all elements of the Mueller matrix. The aluminium mirror acts on a $\sim10^{-3}$ level as a polarizer due to diattenuation (elements $I \Rightarrow Q$ and $Q \Rightarrow I$) and as a quarter wave plate due to retardance (elements $U \Rightarrow V$ and $V \Rightarrow U$). The structure in the upper right and lower left 2x2 submatrices in the PSM and Mueller pupil originate from modal crosstalk effects. The structure in the $V \Rightarrow I$ Mueller pupil element (\autoref{fig:Mueller_Pupil}) is on a level below the current numerical precision of the code for absorbing materials (see \autoref{fig:Snells_Law} and \autoref{fig:Fresnel_Coefficients_Glass_Air}) and can therefore not be trusted. \\

\section{Conclusion}
We have started the development of a code to simulate optical systems and characterize geometrical and polarization aberrations. The code uses rigorous vector diffraction in homogeneous materials to achieve high precision ($\sim10^{-15}-10^{-16}$) for non-absorbing dielectric materials. For absorbing materials the precision is on the order of $\sim10^{-8}$, which can be further improved. Current features are interference effects in thin films and unpolarized light.\\
\indent Future work includes fully integrating rotated planes into the framework of the code, propagation to arbitrarily shaped surfaces and verification of the code by comparing it to existing codes such as Polaris-M, Code V and lab experiments.

\bibliography{report}   

\begin{thebibliography}{10}

\bibitem{melville2011computational}
D.~O. Melville, A.~E. Rosenbluth, A.~Waechter, {\em et~al.}, ``Computational
  lithography: exhausting the resolution limits of 193-nm projection
  lithography systems,'' {\em Journal of Vacuum Science \& Technology B,
  Nanotechnology and Microelectronics: Materials, Processing, Measurement, and
  Phenomena} {\bf 29}(6), 06FH04  (2011).

\bibitem{hansen1988overcoming}
E.~W. Hansen, ``Overcoming polarization aberrations in microscopy,'' in {\em
  1988 Los Angeles Symposium--OE/LASE'88},  190--203, International Society for
  Optics and Photonics  (1988).

\bibitem{breckinridge2015polarization}
J.~B. Breckinridge, W.~S.~T. Lam, and R.~A. Chipman, ``Polarization aberrations
  in astronomical telescopes: the point spread function,'' {\em Publications of
  the Astronomical Society of the Pacific} {\bf 127}(951), 445  (2015).

\bibitem{chipman2015polaris}
R.~A. Chipman and W.~S.~T. Lam, ``The polaris-m ray tracing program,'' in {\em
  SPIE Optical Engineering+ Applications},  96130J--96130J, International
  Society for Optics and Photonics  (2015).

\bibitem{ruoff2009orientation}
J.~Ruoff and M.~Totzeck, ``Orientation zernike polynomials: a useful way to
  describe the polarization effects of optical imaging systems,'' {\em Journal
  of Micro/Nanolithography, MEMS, and MOEMS} {\bf 8}(3), 031404--031404
  (2009).

\bibitem{mcleod2014vector}
R.~R. McLeod and K.~H. Wagner, ``Vector fourier optics of anisotropic
  materials,'' {\em Advances in Optics and Photonics} {\bf 6}(4), 368--412
  (2014).

\bibitem{francis1961qr}
J.~G. Francis, ``The qr transformation a unitary analogue to the lr
  transformation: Part 1,'' {\em The Computer Journal} {\bf 4}(3), 265--271
  (1961).

\bibitem{francis1962qr}
J.~G. Francis, ``The qr transformation: Part 2,'' {\em The Computer Journal}
  {\bf 4}(4), 332--345  (1962).

\bibitem{kublanovskaya1962some}
V.~N. Kublanovskaya, ``On some algorithms for the solution of the complete
  eigenvalue problem,'' {\em USSR Computational Mathematics and Mathematical
  Physics} {\bf 1}(3), 637--657  (1962).

\bibitem{born2013principles}
M.~Born and E.~Wolf, {\em Principles of optics: electromagnetic theory of
  propagation, interference and diffraction of light}, Elsevier  (2013).

\bibitem{semel2003spectropolarimetry}
M.~Semel, ``Spectropolarimetry and polarization-dependent fringes,'' {\em
  Astronomy \& Astrophysics} {\bf 401}(1), 1--14  (2003).

\bibitem{lavrinenko2015numerical}
A.~V. Lavrinenko, J.~L{\ae}gsgaard, N.~Gregersen, {\em et~al.}, {\em Numerical
  Methods in Photonics}, CRC Press  (2015).

\bibitem{li1996formulation}
L.~Li, ``Formulation and comparison of two recursive matrix algorithms for
  modeling layered diffraction gratings,'' {\em JOSA A} {\bf 13}(5), 1024--1035
   (1996).

\bibitem{berreman1972optics}
D.~W. Berreman, ``Optics in stratified and anisotropic media: 4$\times$
  4-matrix formulation,'' {\em Josa} {\bf 62}(4), 502--510  (1972).

\bibitem{mccall2014birefringent}
M.~W. McCall, I.~J. Hodgkinson, and Q.~Wu, {\em Birefringent thin films and
  polarizing elements}, World Scientific  (2014).

\bibitem{weenink2012application}
J.~Weenink, F.~Snik, C.~Keller, {\em et~al.}, ``Application of the berreman
  calculus to polarized spectral fringes in wave plates,'' {\em A\&A (to be
  submitted)}   (2012).

\bibitem{mcguire1990diffraction}
J.~P. McGuire and R.~A. Chipman, ``Diffraction image formation in optical
  systems with polarization aberrations. i: Formulation and example,'' {\em
  JOSA A} {\bf 7}(9), 1614--1626  (1990).

\bibitem{soummer2007fast}
R.~Soummer, L.~Pueyo, A.~Sivaramakrishnan, {\em et~al.}, ``Fast computation of
  lyot-style coronagraph propagation,'' {\em Optics Express} {\bf 15}(24),
  15935--15951  (2007).

\bibitem{wolf2007introduction}
E.~Wolf, {\em Introduction to the Theory of Coherence and Polarization of
  Light}, Cambridge University Press  (2007).

\bibitem{bass2009handbook}
M.~Bass, C.~DeCusatis, J.~Enoch, {\em et~al.}, {\em Handbook of optics, Volume
  II: Design, fabrication and testing, sources and detectors, radiometry and
  photometry}, McGraw-Hill, Inc.  (2009).

\bibitem{bindel2002computing}
D.~Bindel, J.~Demmel, W.~Kahan, {\em et~al.}, ``On computing givens rotations
  reliably and efficiently,'' {\em ACM Transactions on Mathematical Software
  (TOMS)} {\bf 28}(2), 206--238  (2002).

\bibitem{wang2002convergence}
T.-L. Wang and W.~Gragg, ``Convergence of the shifted ???? algorithm for
  unitary hessenberg matrices,'' {\em Mathematics of computation} {\bf
  71}(240), 1473--1496  (2002).

\bibitem{aiello2009duality}
A.~Aiello, M.~Merano, and J.~Woerdman, ``Duality between spatial and angular
  shift in optical reflection,'' {\em Physical Review A} {\bf 80}(6), 061801
  (2009).

\bibitem{lvovsky2013fresnel}
A.~I. Lvovsky, ``Fresnel equations,'' {\em Encyclopedia of Optical Engineering;
  Taylor \& Francis: New York, NY, USA} , 1--6  (2013).

\bibitem{merano2007observation}
M.~Merano, A.~Aiello, M.~Van~Exter, {\em et~al.}, ``Observation of
  goos-h{\"a}nchen shifts in metallic reflection,'' {\em Optics express} {\bf
  15}(24), 15928--15934  (2007).

\bibitem{hecht2002optics}
E.~Hecht, ``Optics, 4th,'' {\em International edition, Addison-Wesley, San
  Francisco} {\bf 3}  (2002).

\bibitem{mcpeak2015plasmonic}
K.~M. McPeak, S.~V. Jayanti, S.~J. Kress, {\em et~al.}, ``Plasmonic films can
  easily be better: Rules and recipes,'' {\em ACS photonics} {\bf 2}(3),
  326--333  (2015).

\end{thebibliography}
\bibliographystyle{spiejour}   

\end{document}